%% file: disc2011.tex
\documentclass{llncs}
\usepackage{epsfig}
\usepackage{textcomp}
\usepackage[latin1]{inputenc}
\usepackage{amssymb}
\usepackage{listings}
\usepackage{epsf,amsmath}
\usepackage{subfig}
\usepackage{graphics}
\usepackage{multirow} 
\usepackage{color}
\definecolor{orange}{rgb}{1,0.5,0}

\newcommand{\ie}{{i.\,e.}}

\renewcommand{\epsilon}{\varepsilon}

\begin{document}
\sloppy

\title{Light-weight Locks\vspace{-3mm}}

\author{Nitin Garg\inst{1} \and Ed Zhu\inst{1} \and Fabiano C. Botelho\inst{1}}

\institute{
Data Domain, an EMC$^2$ Company,
Santa Clara, CA, USA \\
\texttt{\{nitin.garg, ed.zhu, fabiano.botelho\}@emc.com}
}

\maketitle

\begin{abstract}
\vspace{-5mm}
In this paper, we propose a new approach to building synchronization primitives,
dubbed ``lwlocks'' (short for light-weight locks). The primitives are optimized
for small memory footprint while maintaining efficient performance in low
contention scenarios. A read-write lwlock occupies 4 bytes, a mutex occupies
4 bytes (2 if deadlock detection is not required), and a condition variable
occupies 4 bytes. The corresponding primitives of the popular pthread library
occupy 56 bytes, 40 bytes and 48 bytes respectively on the x86-64 platform. The
API for lwlocks is similar to that of the pthread library but covering only the most
common use cases. Lwlocks allow explicit
control of queuing and scheduling decisions in contention situations and support
``asynchronous'' or ``deferred blocking'' acquisition of locks. Asynchronous locking
helps in working around the constraints of lock-ordering which otherwise limits concurrency.
The small footprint of lwlocks enables the construction of data structures with 
very fine-grained locking, which in turn is crucial for lowering contention and supporting highly concurrent
access to a data structure. Currently, the Data Domain File System uses lwlocks for its 
in-memory inode cache as well as in a generic doubly-linked concurrent list which forms
the building block for more sophisticated structures.
\vspace{-2mm}
\end{abstract}
\sloppy
\lstset{
  language=C,
  basicstyle=\fontsize{8}{8}\selectfont,
   aboveskip=0mm,
   belowskip=0mm,
   mathescape=true,
   escapechar=@,
   showstringspaces=false,
   columns=fixed,
   basewidth=0.515em,
   frame=none,
   framesep=1mm,
}

\input{introduction}

\input{idea}

\input{how_it_works}

\input{q_manage_and_async_locks}

\input{performance}

\input{conclusions}


\bibliographystyle{plain}

\bibliography{disc2011}
\clearpage
\appendix
\input{appendix}

\end{document}

%% file: introduction.tex
\section{Introduction}
\label{introduction}
The advent of the multi-core systems has forced a rethinking of basic data structures in order to support greater scalability and concurrency~\cite{s11}. While there have been good strides in building lock-free versions of certain data structures~\cite{K03,ls04}, and software transactional memory (STM) based techniques are becoming popular~\cite{rrhw09,st97}, the use of traditional locking techniques remains the de-facto standard for synchronization in shared-memory systems. The usual technique for increasing concurrency using traditional locking schemes, aside from using algorithms that reduce the concurrent sections~\cite{KB95,LY81}, is to use different locks for different parts of the data structures. The use of such fine-grained locking often runs afoul of the overhead involved, thereby limiting the maximum number of locks used. To minimize the space overhead, the algorithms usually try to minimize the number of locks, and in turn need to build a mapping to and from different parts of the structure to the corresponding lock. This adds to the complexity of the code that needs to be maintained.

In this paper we present a novel technique to create locking primitives that have a very small memory footprint. We call our locks ``light-weight locks'' or ``lwlocks''. Specifically, a read-write lock in our scheme takes 4 bytes, a mutex takes 4 bytes (only 2 if deadlock detection is not required), and a condition variable takes 4 bytes. The corresponding primitives of the popular pthread library 
occupy 56 bytes, 40 bytes and 48 bytes respectively on the x86-64 platform. The API for lwlocks is modeled after that of the pthread library. We however eschew some of the features provided by pthread locks for the sake of simplicity of our implementation. 
 
We consider our contributions as being four-fold:
(i) locking primitives with small memory footprint which makes them ideal for very fine-grained locking;
(ii) the mechanism underlying the implementation of lwlocks that allows creation of custom lock-like primitives;
(iii) access to waiting queue of threads so custom scheduling schemes can be implemented; and
(iv) support for ``asynchronous'' or ``deferred block'' locking.


In this paper, we focus largely on lwlocks. The rest of the paper is organized as follows: Section~\ref{idea} describes the idea that forms the basis of lwlocks. Section~\ref{how_it_works} describes the internal structure of the supported primitives and the algorithms for implementing their APIs. Section~\ref{q_manage_and_async_locks} briefly describes possible extensions to lwlocks and how asynchronous locking works. Section~\ref{performance} compares the performance of lwlocks with the corresponding primitives in the pthread library. Finally, in Section~\ref{conclusion} we present our conclusions.

%% file: idea.tex
\section{The Fundamental Idea}
\label{idea}

The core idea behind lwlocks is the observation that while a thread could block on different locks or wait on many different condition variables in its lifetime, it can block on only one lock or condition variable at any given point. With lwlocks, whenever a thread has to block, it uses a ``waiter'' structure to do so. In this paper, we use the term ``waiter structure'' or simply ``waiter'' interchangeably. 
Each thread has its own waiter structure and can access it by invoking the \texttt{tls\_get\_waiter} function (which returns the pointer to the waiter kept in the thread local storage).

Figure~\ref{tls} presents the definition of a waiter structure. For compact representation, we limit the maximum number of waiter structures to be less than $2^{16}$ so that each structure can also be uniquely referred to by a 16-bit number. We reserve the value $2^{16} - 1$ to represent the \texttt{NULL} waiter structure and denote it by \texttt{NULLID}. We expect the limit on number of waiters, and hence on the number of threads, to be large enough for most applications\footnote{The limit can be increased for a small increase in the size of the locks which presumably will be acceptable for an application that can support so many threads.}.
\begin{figure}
\vspace{-2mm}
\begin{center}
\begin{lstlisting}[multicols=2]
@{\em // The following definitions assumes }@
@{\em // that each bool\_t takes 4 bytes, each }@
@{\em // pthread\_mutex\_t takes 40 bytes, each}@ 
@{\em // pthread\_cond\_t takes 48 bytes, and}@ 
@{\em // each function pointer takes 8 bytes.}@
@{\bf interface}@ event_t {
  void signal();
  void wait();
  bool_t poll();
} @{\em // 24 bytes}@

@{\bf interface}@ domain_t {
  waiter_t alloc_waiter();
  void free_waiter(waiter_t waiter);
  waiter_t get_waiter();
  waiter_t id2waiter(uint16_t id);
} @{\em // 32 bytes}@

struct waiter_t {
  event_t  event;
  domain_t domain;
  bool_t signal_pending;
  bool_t waiter_waiting;
  pthread_mutex_t mutex; 
  pthread_cond_t  cond; 
  uint64_t app_data;
  uint16_t id;
  uint16_t next;
  uint16_t prev;
} @{\em // 166 bytes}@
\end{lstlisting}
\end{center}
\vspace{-3mm}
\caption{Definition of a waiter structure.}
\label{tls}
\vspace{-3mm}
\end{figure}

A waiter structure is assigned to a thread the first time the thread accesses it (via \texttt{tls\_get\_waiter}) and the structure is returned to the pool of free waiter structures when the thread exits to be re-used by a later thread. The waiter structure is the key piece that enables the compact nature of lwlocks. It can also be used to create other custom compact lock-like data structures. The current non-optimized implementation of a waiter structure occupies $166$ bytes. Since this cost is per thread and we expect the normal use case to have far fewer threads than the number of locks, the amortized cost is very low. For example, an application with $1,024$ threads and around $4,725$ mutexes will have the same memory footprint whether using lwlocks or pthread mutexes. Our normal expected use case is for applications that need several hundred thousands or millions of locks which would make lwlocks the clear choice for locking. 
We now describe the most important pieces that form a waiter structure.

{\bf Waiter's Event.}
A generic event interface underlies the actual mechanics that are used by a thread when it blocks or unblocks on a lock. The two main operations defined for an event are: (i) \texttt{wait}, which is called to wait for the event to trigger; and (ii) \texttt{signal}, which informs a waiter of an event getting triggered. A waiter structure uses one pthread mutex and one condition variable to implement both operations. The operation \texttt{wait} blocks the thread on the condition variable until a signal arrives. The operation \texttt{signal} wakes up the blocked thread. Like semaphores, the implementation ensures that a signal on an event cannot be lost, \ie, a \texttt{signal} can be invoked before the matching \texttt{wait} is and the \texttt{wait} will find the pending signal. Unlike a semaphore, however, the operations \texttt{wait} and \texttt{signal} are always called in pairs. There is also a third operation called \texttt{poll}. It can be used to check if a signal is already pending.

{\bf Waiter's Domain.}
Instead of a fixed implementation for mapping from an id to the waiter structure, we have abstracted out the notion of a waiter's domain. A waiter's domain defines four operations: (i) \texttt{alloc\_waiter} to allocate a waiter from the domain; (ii) \texttt{free\_waiter} to return a waiter back to the domain; (iii) \texttt{get\_waiter} which allows a thread to get to its own waiter; and (iv) \texttt{id2waiter} to map from an id to the waiter structure.

Abstracting the notion of a domain has three benefits. First it provides one more way of extending the system so that instead of an entire application being limited to a maximum of $2^{16}$ threads, the limit only applies to individual domains. Second it provides the flexibility to create domains that have lower limit on maximum concurrency, thereby allowing for creation of locks with even smaller footprint. For example, a system limiting itself to 15 threads (127 without deadlock detection) would need only 1 byte for a mutex. 
Third, combining it with a custom event, allows for creation of libraries such as a user space job scheduler. The \texttt{wait} call on a job blocks it and causes the scheduler to switch to another ready job while the \texttt{signal} call marks the job as ready again. We mention the waiter's domains only for completeness as they are not necessary to understand the workings of lwlocks. Lwlocks use a default global domain whose waiter structures implement the behavior we describe here.

{\bf Forming Lists or Stacks of Waiters.}
Each waiter structure records its own id. It also has space for previous and next id values which can be used to form stacks or lists of waiters. Such a list (or stack) of waiters can be identified purely by the id of the first element of the list, \ie, it can be represented by a 16-bit value. To go to the next (previous) waiter, we convert the current id to the corresponding waiter structure and look at the next (previous) id field in it. 

{\bf Locking Data.}
The final important piece of a waiter structure is the space it provides that can be used by the abstractions built on top for their own purpose. The waiter itself does not interpret it in any way. For instance, read-write lwlocks use this space to record the type of locking operation that the thread was performing when it blocked: whether it was taking a read or write lock. Currently, this space amounts to 8 bytes and is referred to as \texttt{app\_data}.

%% file: how_it_works.tex
\section{Light-Weight Lock Primitives}
\label{how_it_works}

We now look at the internals of each of the lwlock primitives, the supported operations and how they work. The lwlocks by default are ``fair'': a lock is acquired in FIFO order by the threads blocked on it and wake-ups from a condition variable are done in the order in which the threads called \texttt{wait} on the condition variable. Pthread locks are not fair in this sense, and although it is possible to build lwlocks to mimic the same behavior, we have found fairness to be better suited to our needs in the Data Domain File System~\cite{zlp08}. 

Each primitive uses 2 bytes to keep a queue of waiter structures of the threads that are blocked on that primitive. This queue is aptly called a \texttt{waitq}. The \texttt{waitq} is maintained as a ``reverse list'' as that allows insertion of a new waiter in a single hardware supported compare-and-swap (CAS) instruction. The \texttt{next} field of a waiter structure holds the id of the waiter structure in front of it. The oldest waiter's \texttt{next} field holds \texttt{NULLID}. The oldest waiter is the waiter in front of the \texttt{waitq}. 

To acquire a lock, a thread uses the CAS instruction to either take ownership of the lock or add its own waiter structure to the lock's \texttt{waitq}. If the lock is acquired, nothing more needs to be done. If it cannot be acquired, then the thread waits on its waiter structure (by calling the event's \texttt{wait} routine). When the thread's turn comes to own the lock (in FIFO order), the unlocking thread will transfer the lock to it and invoke the event's \texttt{signal} routine on the waiter to wake up the thread. Since the unlocking thread does the work of transferring the lock state and ownership, the waking thread can assume that it has the lock upon being signaled\footnote{For unfair locks, this part has to change and the waking thread would need to try again to take the lock.}. The unlocking thread has to walk the \texttt{waitq} to find the waiter to signal. At any point there can be only one thread performing the transfer on a lock and hence the walk is safe to perform. 

We now present each one of the lwlock primitives. Note that we only highlight the essence of the various operations in the included algorithms. The actual implementation, which we hope to release to the open source community in the near future, has additional logic for performance optimization.

{\bf Light-weight mutex.}
The 4 bytes of a light-weight mutex (henceforth a \texttt{lwmutex}) are composed of 2 equal parts. The first part holds the id of the waiter structure of the owner thread and the second part is the \texttt{waitq}. The owner id is necessary to do self-deadlock detection. Figure~\ref{lwmutex_algos} outlines the lock and unlock algorithms for the 4-byte version of a \texttt{lwmutex}. If deadlock detection is not required, the lock only needs to be 16 bits in size to hold the \texttt{waitq}. To comply with POSIX semantics, we also need to be able to ascertain the owner of such a mutex. Fortunately, we can use the same \texttt{waitq} space. The locking thread swaps the \texttt{NULLID} of the \texttt{waitq} with the id of its own waiter to indicate that the lock is taken. As other threads block, their waiter structures get added to the \texttt{waitq} as in the case of the regular \texttt{lwmutex}. The difference is that the \texttt{next} field of the waiter structure in front of the \texttt{waitq} does not hold \texttt{NULLID}. Instead, it holds the id of the waiter of the lock owner thread. Hence, the unlock operation traverses the \texttt{waitq} until a waiter whose \texttt{next} field matches the id of the unlocking thread's waiter structure is reached. The \texttt{next} field is reset to \texttt{NULLID} and the waiter is signaled.



\begin{figure}
\vspace{-3mm}
\begin{center}
\begin{lstlisting}[multicols=2]
struct lwmutex_t {
  uint16_t owner; @{\em // owner ID }@
  uint16_t waitq; @{\em // tail of queue}@
}

void lock(lwmutex_t @$m$@) {
  @$w$@ = tls_get_waiter();
  do {
    @$n$@ = @$o$@ = @$m$@;
    if (@$o$@.owner @==@ @NULLID@)
      @$n$@.owner = @$w$@.id;
    else {
      @$w$@.next = @$n$@.waitq;
      @$n$@.waitq = @$w$@.id;
    }
  } while (!CAS(@$m$@, @$o$@, @$n$@));
  if (@$n$@.owner @==@ @$w$@.id)
    return; // Got lock
  // Wait for lock transfer
  @$w$@.event.wait();
}

void unlock(lwmutex_t @$m$@) {
  @$w$@ = tls_get_waiter();    
  do {
    @$n$@ = @$o$@ = @$m$@;
    if (@$o$@.waitq != @NULLID@) {
      @$wtw$@ = id2waiter(@$m$@.waitq);
      @$pw$@ = @NULL@;
      while (@$wtw$@.next != @NULLID@) {
        @$pw$@ = @$wtw$@;
        @$wtw$@ = id2waiter(@$wtw$@.next);
      }
      if (@$pw$@ == @NULL@) @$n$@.waitq = @NULLID@;
      @{\em // Transfer lock to $wtw$}@   
      @$n$@.owner = @$wtw$@.id;
    } else @$n$@.owner = @NULLID@;
  } while (!CAS(@$m$@, @$o$@, @$n$@));
  // @{\em remove}@ @$wtw$@ @{\em from q}@
  if (@$pw$@ != @NULL@) 
     @$pw$@.next = @NULLID@; 
  if (@$wtw$@) @$wtw$@.event.signal();
}
\end{lstlisting}
\end{center}
\vspace{-5mm}
\caption{Operations to lock and unlock a \texttt{lwmutex}.
The old and new values passed in to CAS are denoted by $o$ and $n$,
respectively. The caller's thread local waiter structure is denoted by $w$.
We use $wtw$ and $pw$ to denote the waiter to wake up and the 
previous waiter in the queue, respectively.
}
\vspace{-3mm}
\label{lwmutex_algos}
\end{figure}
    
{\bf Light-weight condition variable.}
%
The 4 bytes of a light-weight condition variable (henceforth a \texttt{lwcondvar}) are composed of 2 equal parts. The first part is a 2-byte version of \texttt{lwmutex} and the second part is the queue of waiter structures. There are three basic operations for a \texttt{lwcondvar}: (i) \texttt{wait}; (ii) \texttt{signal}; and (iii) \texttt{broadcast}. The internal 2-byte \texttt{lwmutex} is used to synchronize manipulation of the waiter's queue which makes the algorithms for those three operations very easy to derive. The algorithms for the three operations are presented in Appendix~\ref{pseudo_lwcondvar}.


{\bf Light-weight read-write lock.}
%
The light-weight read-write lock (henceforth a \texttt{lw\_rwlock}) also uses 2 of its 4 bytes for the \texttt{waitq}. Of the remaining 16 bits, 14 bits are used for the count of read locks granted, 1 bit is used to indicate a write lock, and the final 1 bit is used to indicate whether the lock is read-biased or not. A read-biased lock is unfair towards writers in the sense that a thread that needs a read lock will acquire it without any regard to waiting writers if the lock is already held by other readers. This behavior is similar to that of pthread read-write lock and is essential for applications where a thread can recursively acquire the same lock as a reader. Without the read-biased behavior, a deadlock can result if a writer arrives in between two read lock acquisitions: the second read lock attempt will wait for the writer which is waiting for the first read lock to be released. Applications that do not have recursive read locking do not need the read-biased behavior but may choose to use it for throughput reasons.

The 14-bit reader count limits the maximum number of readers per lock to $2^{14}$, a limit that we have found to be sufficient in practice. The limit can be raised by having the API explicitly flag read-bias behavior, so the bias bit does not have to be in the \texttt{lw\_rwlock} or restricting the maximum concurrency, thereby freeing bits from the \texttt{waitq} or by slightly increasing the size of the lock. 

Figure~\ref{rwlock_algos} outlines the algorithms for the two main operations: (i) \texttt{lock}, and (ii) \texttt{unlock}. The lock operation on \texttt{lw\_rwlock} is similar to that of \texttt{lwmutex} with the added flag indicating if a read or write lock is requested. The unlock operation for non-read-biased \texttt{lw\_rwlock} has to pick the oldest set of waiters that it can signal: either a single writer or a set of contiguous readers. A read-biased \texttt{lw\_rwlock} can follow the same logic as a non-biased \texttt{lw\_rwlock} when the transfer is to a waiting writer. For the transfer from a writer to reader(s), however, the writer has to signal all readers, not just the oldest contiguous set. The solution is to have the writer atomically remove the entire \texttt{waitq} and downgrade to a read lock. It then separates the \texttt{waitq} into two queues: one consisting of readers and one consisting of writers. The writers are added back to the front of the \texttt{waitq} while also updating the reader count to fully account for the readers found in the removed \texttt{waitq}. Finally, the readers can be signaled. Note that the re-insert of the waiting writers during unlock is safe. The re-insert is done at the front of the \texttt{waitq} and any new writers will add themselves to the back of the \texttt{waitq}. No other thread can be traversing the \texttt{waitq} for ownership transfer as the re-inserting thread holds a read lock. This case makes the implementation of \texttt{lw\_rwlocks} the most complex of all the primitives and the algorithm outline is only at a high level for the contention case, where the \texttt{waitq} has at least one waiter in it. The non-contented case is simple to derive.

\begin{figure}
\vspace{-3mm}
\begin{center}
\begin{lstlisting}[multicols=2]
struct lw_rwlock_t {
  uint1_t rd_bias;  
  uint1_t wlocked;  
  uint14_t readers; 
  uint16_t waitq; 
}
void @lock(lw\_rwlock\_t $\ell$, bool\_t exclusive)@ {
  @$w$@ = tls_get_waiter();
  do {
    @$o$@ = @$n$@ = @$\ell$@;
    if (!exclusive && !@$o$@.wlocked && 
       (@$o$@.waitq @==@ @NULLID@ || 
        @$o$@.rd_bias))
      @$n$@.readers++;
    else if (@exclusive@ && 
            !(@$n$@.wlocked || @$n$@.readers > 0))
      @$n$@.wlocked = 1;
    else { @{\em // Need to block}@
      @$w$@.app_data = exclusive;
      @$w$@.next = @$o$@.waitq;
      @$n$@.waitq = @$w$@.id;
    }
  } while (!CAS(@$\ell$@, @$o$@, @$n$@));
  if @($n$.waitq == $w$.id) $w$.event.wait();@
}
void @unlock\_fair(lw\_rwlock\_t $\ell$)@ {
  do {
    @$o$@ = @$n$@ = @$\ell$@;
    if (@$n$@.wlocked == 1) @$n$@.wlocked = 0;
    else @$n$@.readers--;
    if(!(@$n$@.wlocked || @$n$@.readers > 0)) { 
      @$(pw, wtw)$ = find\_oldest\_set\_of\_waiters($n$)@;
      if (@$pw$@ == @NULL@) @$n$@.waitq = @NULLID@;
      if (@$wtw$@.app_data != exclusive) { 
        @$n$@.readers = waitq_size(@$wtw$.id@);
      } else @{\em // single writer picked}@
         @$n$@.wlocked = 1;
    }    
  } while (!CAS(@$\ell$@, @$o$@, @$n$@));
  if (@$pw$@ != @NULL@) @$pw$@.next = @NULLID@;
  wake_up_waiters(@$wtw$@);
}
uint16_t waitq_size(uint16_t @$wid$@) {
  uint16_t count = 0;
  while (@$wid$@ != @NULLID@) {
    @$wid$@ = id2waiter(@$wid$@).next;
    count++;
  }
  return count;
}
void @unlock(lw\_rwlock\_t $\ell$)@ {
  if (!@$\ell$@.rd_bias)
    return unlock_fair($\ell$);
  if ($\ell$.wlocked == 0) {
    do { @{\em // Only writers in waitq}@
      @$o$@ = @$n$@ = @$\ell$@;
      if (@$n$@.readers == 1) {
        @$n$@.wlocked = 1; @$n$@.readers = 0;
      } else @$n$@.readers--;
    } while (!CAS(@$\ell$@, @$o$@, @$n$@));
    if (@$n$@.wlocked) unlock_fair($\ell$);
    return;
  }
  @{\em // writer unlocking a biased lock}@
  @$ow$@ = find_oldest_waiter(@$\ell$@); 
  if (@$ow$@.app_data == exclusive) {
    @{\em // handing off to writer}@
    unlock_fair($\ell$); return;
  }
  @{\em // Wake up all readers. Atomically}@
  @{\em // downgrade to read lock and }@
  @{\em // clear \& return waitq. After the}@
  @{\em // downgrade only writers can block on $\ell$.}@
  waitq = downgrade_to_read_lock(@$\ell$@);
  @{\em // split waitq in two subqueues:}@
  @{\em // readers queue and writers queue}@
  (rd_q, wr_q) = splitq(waitq);
  do {
    @$o$@ = @$n$@ = @$\ell$@;
    @{\em // 1 reader added during downgrade.}@
    @$n$@.readers += waitq_size(rd_q) - 1;
    if (@$n$@.waitq == @NULLID@) @$n$@.waitq = wr_q
  } while (!CAS(@$\ell$@, @$o$@, @$n$@));
  if (@$n$@.waitq != wr_q && wr_q != @NULLID@) {
    @$ow$@ = find_oldest_waiter(@$n$@);
    @$ow$@.next = wr_q;
  }
  wake_up_waiters(id2waiter(rd_q));
}
\end{lstlisting}
\end{center}
\vspace{-4mm}
\caption{Operations to lock and unlock a \texttt{lw\_rwlock}. 
The \texttt{lock} operation takes a boolean as input to indicate whether an exclusive lock 
is requested. The old and new values passed in to CAS are denoted 
by $o$ and $n$, respectively. The caller's thread local waiter structure is denoted by $w$.
We use $wtw$, $pw$ and $ow$ to denote the waiter to wake up, the 
previous waiter in the \texttt{waitq}, and the oldest waiter in the \texttt{waitq}, respectively.
}
\vspace{-4mm}
\label{rwlock_algos}
\end{figure}

%% file: q_manage_and_async_locks.tex
\section{Asynchronous Locking and Other Extensions}
\label{q_manage_and_async_locks}

We take a moment here to highlight some aspects of the algorithms presented in Section~\ref{how_it_works} and how small changes would enable alternative behaviors. On the locking side, the key observation is that once a thread has put itself on the wait queue of a lock or condition variable, it is guaranteed to have the lock transferred to it or a signal delivered to it. The thread does not have to call \texttt{wait} right away. The thread could spin for a certain amount of time on \texttt{poll} before calling \texttt{wait} effectively creating adaptive locks. It could also keep spinning which would create starvation-free spin locks. Both of these are scalable and contention-free similar to the approaches in~\cite{A90,GS90,MS91-2}.

Alternately, the \texttt{lock} operation could simply return without calling \texttt{wait} at all. This would allow the calling thread to take some application-specific action before invoking \texttt{wait}. We call this mode of operation as taking an ``asynchronous'' or ``deferred blocking'' lock. Asynchronous locking is the key enabler to work around the constraints that lock-ordering imposes. We use this functionality in building a generic highly concurrent doubly-linked list in the Data Domain File System~\cite{zlp08}. The list allows concurrent appends, dequeues, inserts (before or after any member), deletes and iterators (in either direction). Some of these operations need to acquire locks in opposite order of other operations. To avoid deadlocks, a canonical order is picked and operations that need to acquire locks in the opposite direction use asynchronous locking. 

The following example, taken from doubly-linked list implementation, illustrates how asynchronous locking is used and why it is essential. Suppose the canonical order for nodes A \& B is A, then B. A thread holds a lock on B already and needs to lock A. It will make an asynchornous lock call for A. If the thread is unable to get the lock, it is on A's \texttt{waitq}, and it releases the lock on B. It then waits for the lock on A to be granted and then reacquires the lock on B (which is in canonical order). In the above sequence, the thread always either holds a lock (on A or B) or is in the \texttt{waitq} of a lock (on A). Other guarantees in the data structure assure that in this case A and B will remain valid and hence there will be no illegal access. Achieving this without asynchronous locking is not possible. Using \texttt{trylock} on A and upon failure, releasing B then locking A leaves a window open between release of B and locking of A where neither node is in any way aware of the thread. One or both nodes could go away in that window and the thread would end up performing an illegal access.


We are also working on building highly concurrent versions of other data structures (trees of various kinds) where we expect to use asynchronous locking frequently. Note that since there is only one waiter structure per thread per domain, a thread can only be performing one asynchronous lock operation per domain at any time. To keep the discussion focused on lwlocks, we cannot go into any more details of our list or other data structures here.

On the unlocking side of the operations, we note that since the \texttt{waitq} management is visible in user space, the unlocking thread has a lot of flexibility in picking which of the waiting threads to signal and whether to do lock hand-off or have the signaled thread retry. This can be exploited to create any custom scheduling policy. We could pick the thread with the highest priority or the longest waiting thread or even have applications use the \texttt{app\_data} to define their own preferences. Signaling waiters in LIFO instead of FIFO order would trade fairness for performance as we illustrate in Section~\ref{performance}.

Finally, with most hardware supporting 64-bit CAS instructions, the generic building blocks of 16-bit \texttt{waitq} leaves 48 bits available for building other primitives. For example, we have built semaphore like counters and a combined mutex+condition variable structure, and implemented \texttt{upgrade} and \texttt{downgrade} operations for \texttt{lw\_rwlocks} (see Appendix~\ref{up_down_appendix} for \texttt{lw\_rwlock} algorithms that allow these). Although our implementation has focussed on process-private locks, we believe it is possible to extend the approach to include process-shared locks. For example, the Linux operating system limits the maximum number of processes to $2^{15}$ which would give a natural mapping from the process id to the waiter structure id for the process. The structures could be managed in user space shared memory or the kernel could manage them. Using an actual semaphore would be more appropriate to use in this case to implement the event interface for the waiters.

%% file: performance.tex
\section{Performance}
\label{performance}

We now examine some experimental data to show that the performance of lwlocks is acceptable. The experiments were performed on a 4-socket system with Xeon E7-4860 processors. Each socket has 10 physical (20 with hyper-threading enabled) cores, for a total of 40 (80 hyper-threaded) cores. The machine has 256GB of memory and each core operates at 2.26GHz.

We have carried out three sets of experiments. Each experiment was run 20 times which was enough to get a confidence level of $99\%$ on the presented average values. The first one compares the performance of unfair \texttt{lwmutexes} with unfair pthread mutexes. Unfair mutexes trade off fairness for performance by using the {\em greedy} approach: the unlocking thread can reacquire the lock right away again. This is done to avoid the {\em convoy} problem. We have implemented two versions of unfair \texttt{lwmutexes}: (i) LIFO wake-ups, which wakes up the most recent thread in the \texttt{waitq}; and (ii) FIFO wake-ups, which wakes up the longest-waiting thread in the \texttt{waitq}. The experiment consists of $n$ threads carrying out the same number of operations on a global doubly-linked list protected by a single unfair mutex -- each operation has the same cost. Each thread acquires the global mutex, performs an operation and drops the mutex. There is no activity outside the locked code block except to increment the loop counter.

\begin{figure}[htb]
\centering
\subfloat[Latency per operation on a global list protected by either a single unfair \texttt{lwmutex} (with LIFO wake-ups or FIFO wake-ups), or pthread mutex.]{
\includegraphics[scale=0.48]{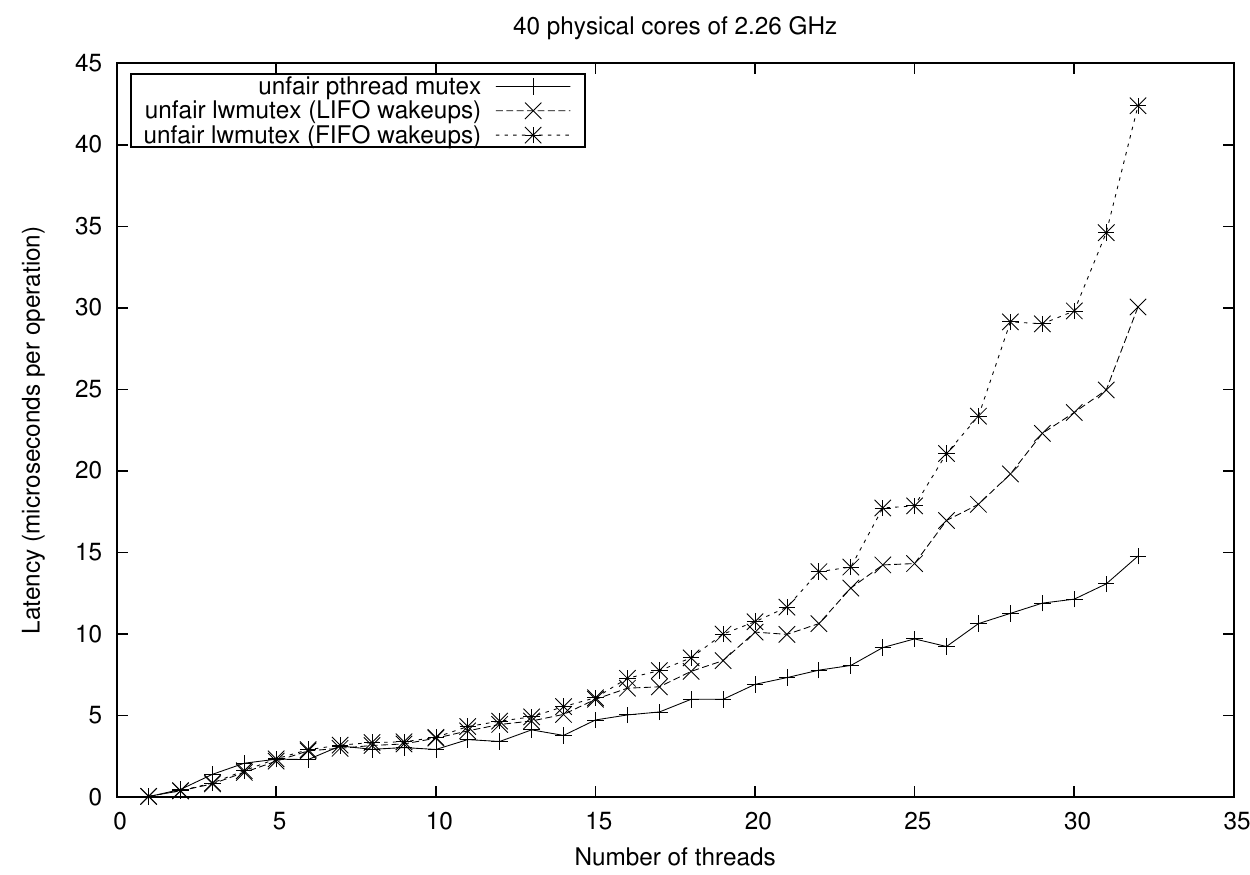}
}\hspace{5mm}
\subfloat[Latency per operation on either a hash table using a fair \texttt{lwmutex} per bucket or $1,024$ unfair pthread mutexes, each one protecting a range of $1,024$ buckets.]{
\includegraphics[scale=0.48]{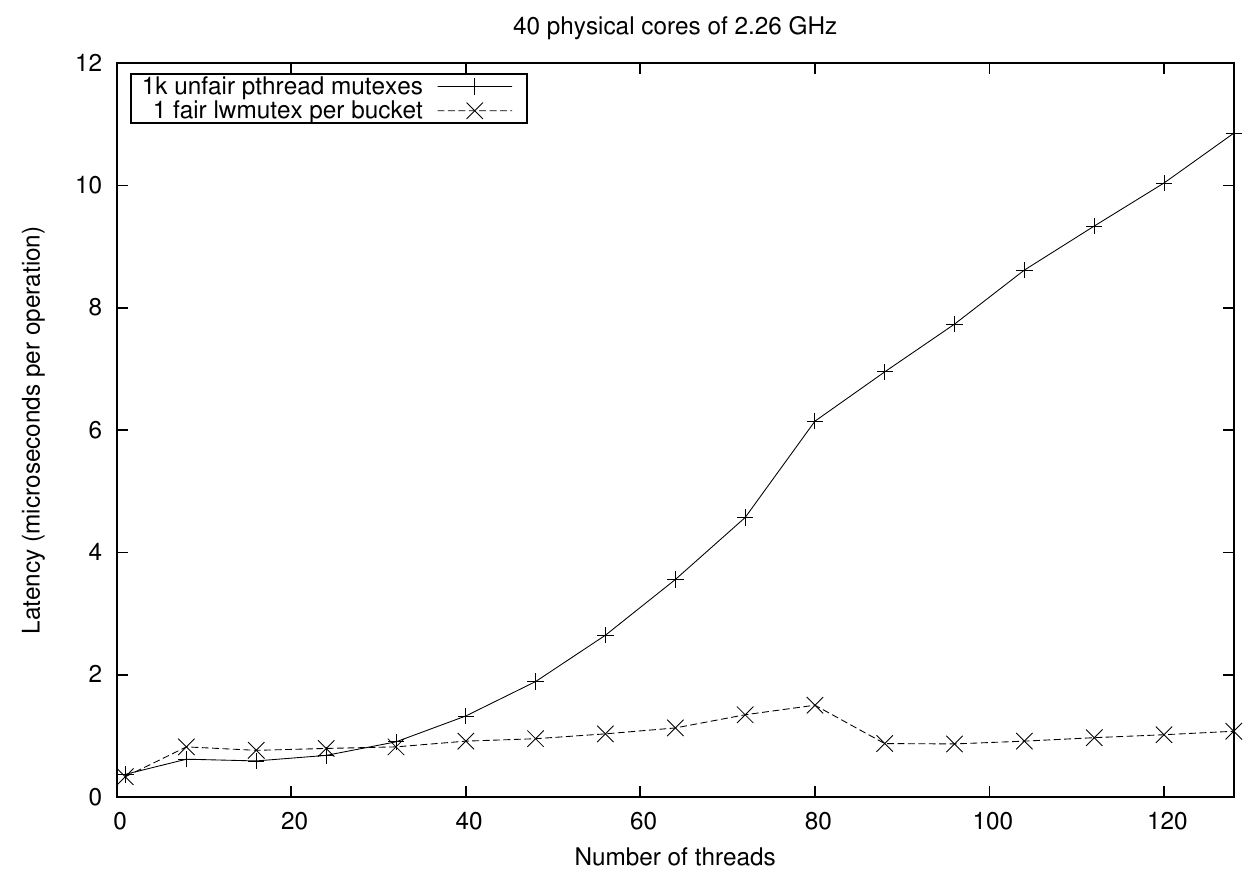}
}
\caption{Latency per operation for: (a) coarse-grained locking; (b) fine-grained locking.}
\label{lwmutexperformance}
\vspace{-2mm}
\end{figure}

Figure~\ref{lwmutexperformance} (a) shows how the latency per operation increases with the number of contending threads. As the number of threads increases, the per operation cost goes up for all lock types. Note that, for relatively low contention ($n \le 10$), unfair \texttt{lwmutexes} perform as good as unfair pthread mutex\footnote{Our code is written entirely in C and compiled with O4 optimization. Pthread code is part C and part fine-tuned assembly.}. We are satisfied that our implementation is reasonably efficient from the performance shown by unfair \texttt{lwmutex}. The gap between pthread mutex and LIFO unfair \texttt{lwmutex} arises from the fact that pthread mutex try the CAS operation only once before making a system call to block. The \texttt{lwmutex} code (both lock and unlock) has to contend until the caller has performed a successful CAS operation. The performance gap betweek LIFO and FIFO version of \texttt{lwmutex} hightlight the overhead of traversing the \texttt{waitq}. It is well known that a fair mutex is considerably slower than an unfair one under high contention due to frequent context switches (the {\em convoy} problem). For 32 contending threads we saw that the latency per operation can go as high as 13x the latency per operation seen for unfair mutexes. However, If there is no contention or just a few contending threads ($< 3$), the latency per operation is very close to the one obtained with unfair mutexes. We note that performance parity with pthread mutexes was never our goal. Although we believe that with proper tuning the cost difference between \texttt{lwmutexes} and pthread locks can be reduced further, our primary concern is the memory overhead that prevents their use in extremely fine-grained locking. Fine-grained locking results in lower contention in general and hence improved performance overall as we show in the next experiment.


The second experiment illustrates how fine-grained locking can deliver better performance overall. The experiment consists of $n$ threads performing lookups, followed by an update to the looked-up record, on a hash table. The hash table has 1 million buckets and is populated with 2 million elements (chaining is used as the collision resolution scheme). We evaluate the latency per operation (in microseconds) for two cases: (i) a fair \texttt{lwmutex} is embedded in each bucket's list head; and (ii) 1,024 unfair pthread mutexes are used, where each one protects a range of 1,024 buckets.

Figure~\ref{lwmutexperformance} (b) shows how the latency per operation increases as the number of threads concurrently operating on the hash table increases. As can be seen, it is preferable to have fine-grained 
locking than optimizing the performance of the lock itself.
Also, when the lock is placed within the bucket itself, it improves the memory locality and may have fewer cache misses compared to accessing pthread mutex located in a separate memory area. For the hash table case is very easy to map from a bucket to a pthread mutex stored in a separate area. That is not true for other data structures like linked lists and trees. Additional logic to minimize the number of locks for those data structures introduces complexity which is more difficult to maintain than for the case where a lock can be cheaply added per node. Even a hash table that uses open-addressing schemes (probing, double hashing or cuckoo hashing~\cite{PR01}) for resolving conflicts presents challenges when using range locking.

Finally, the third experiment compares \texttt{lw\_rwlocks} with read-write pthread locks. We use the same hash table as before but now we fix the number of threads (readers + writers) to 34 and then we vary the number of writers (or contending threads) from 0 to 34. Beyond 34 threads we start seeing contention across readers for pthread locks: the contention is on the update of the reader counter, which is surrounded by a mutex in the pthread library. Because we only want to evaluate the contention due to writers, we in turn, picked 34.

\begin{figure}[htb]
\vspace{-3mm}
\centering
\includegraphics[scale=0.48]{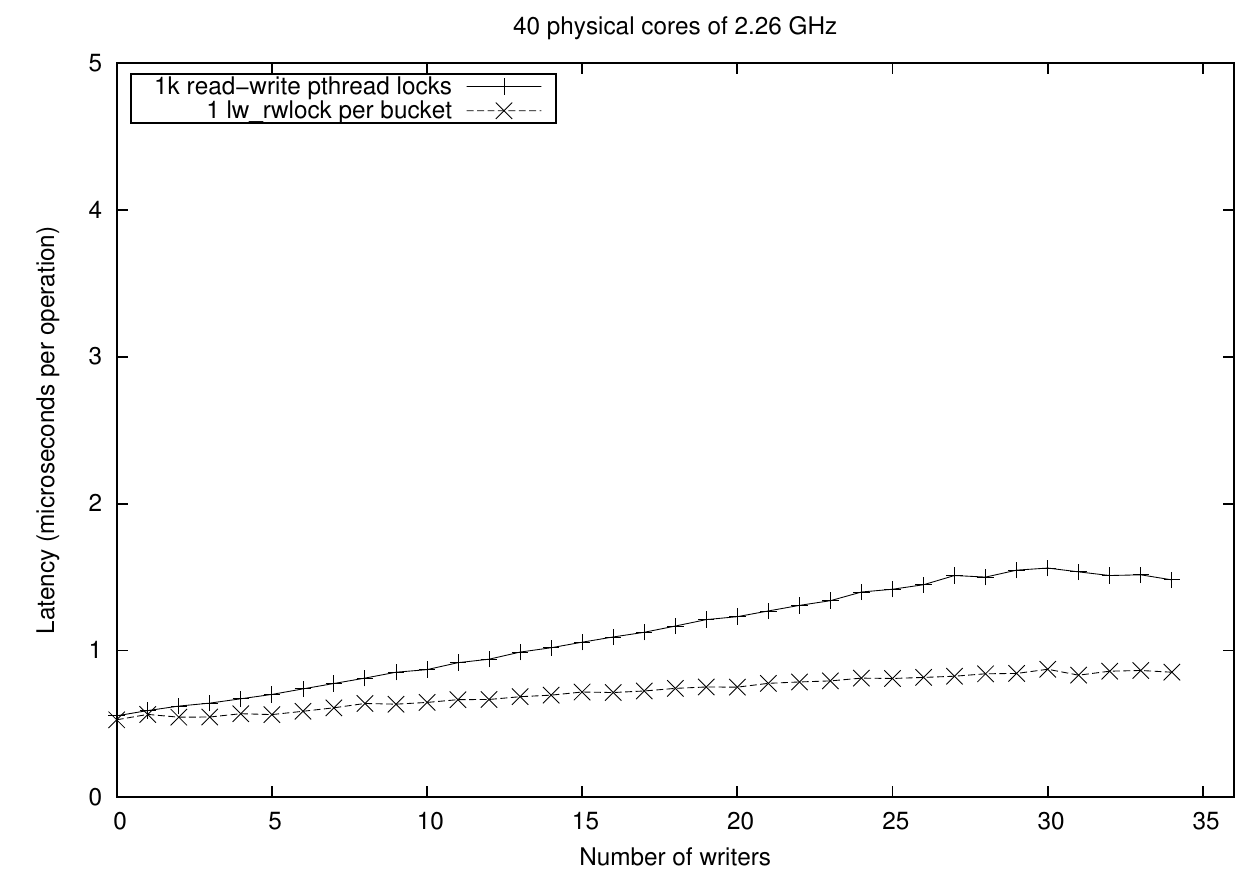}
\vspace{-2mm}
\caption{Latency per operation on either a hash table using a \texttt{lw\_rwlock} per bucket or $1,024$ read-write pthread locks, each one protecting a range of $1,024$ buckets. The total number of threads was fixed to 34 and the number of writers goes from 0 to 34.}
\vspace{-2mm}
\label{lwrwlockperformance}
\end{figure}

Figure \ref{lwrwlockperformance} shows how the latency per operation increases as the number of writers concurrently operating on the hash table increases. Once again the fine-grained locking provided by the cheap \texttt{lw\_rwlocks} delivers better overall performance and also scales better than read-write pthread locks.

%% file: conclusions.tex
\section{Conclusions}
\label{conclusion}

We have presented in this paper a new approach to building compact synchronization primitives. This is possible because each thread can only block in one lock or condition variable at a time. Besides the compact nature of light-weight locks, the queue management of blocked threads is also done entirely in user space. This allows the implementation of features that are impossible to implement with traditional pthread locks. For instance, asynchronous locking cannot be implemented with pthread locks as they stand. The cost for light-weight locks is a 166-byte waiter structure per thread, which amortizes very quickly for applications where there are many more locks than threads. We believe that this is a fairly common scenario.

%% file: appendix.tex
\section{Pseudo code for light-weight condition variables}

Figure~\ref{lwcondvar_algos} presents the structure of a \texttt{lwcondvar} as well as the operations it supports.
\label{pseudo_lwcondvar}
\begin{figure}
\begin{center}
\begin{lstlisting}[multicols=2]
struct lwcondvar_t {
  lwmutex_t m; @{\em // 2-byte lwmutex}@
  uint16_t waitq; @{\em // queue tail}@
}
void @wait(lwmutex\_t $m$@, @lwcondvar\_t $c$)@ {
  @$w$@ = tls_get_waiter();
  @$w$@.next = @NULLID@;
  lock(@$c$@.@$m$@);
  if (@$c$@.waitq == @NULLID@)
    @$c$@.waitq = @$w$@.id;
  else {
    @$w$@.next = @$c$@.waitq;
    @$c$@.waitq = @$w$@.id;
  }
  unlock(@$c$@.@$m$@);
  unlock(@$m$@);
  @$w$@.event.wait(); 
  lock(@$m$@);
}

void wake_up_waiters(waiter_t @$w$@) {
  while (@$w$@ != @NULL@) {
      waitq = @$w$@.next;
      @$w$@.event.signal();
      @$w$@ = id2waiter(waitq) 
  }
}

void @signal(lwcondvar\_t $c$)@ {
  lock(@$c$@.@$m$@);
  if (@$c$@.waitq != @NULLID@) {
    @$wtw$@ = id2waiter(@$c$@.waitq);
    @$pw$@ = @NULL@;
    while (@$wtw$@.next != @NULLID@) {
      @$pw$@ = @$wtw$@;
      @$wtw$@ = id2waiter(@$wtw$@.next);
    }
    if (@$pw$@ == @NULL@) @$c$@.waitq = @NULLID@;
    else @$pw$@.next = @NULLID@;
  } else @$wtw$@ = @NULL@;
  unlock(@$c$@.@$m$@);
  if (@$wtw$@ != @NULL@) @$wtw$@.event.signal(); 
  @{\em // Else missed signal}@
}    

void @broadcast(lwcondvar\_t $c$)@ {
  lock(@$c$@.@$m$@);
  waitq = @$c$@.waitq; @{\em // pointer to queue's head}@ 
  @$c$@.waitq = @NULLID@;
  unlock(@$c$@.@$m$@);
  wake_up_waiters(id2waiter(waitq));
}
\end{lstlisting}
\end{center}
\caption{Operations defined for a \texttt{lwcondvar}. 
The old and new values passed in to CAS are denoted by $o$ and $n$,
respectively. The caller's thread local waiter structure is denoted by $w$.
We use $wtw$ and $pw$ to denote the waiter to wake up and the 
previous waiter in the queue, respectively.
}
\label{lwcondvar_algos}
\end{figure}

\section{Upgrading and Downgrading light-weight read-write locks}
\label{up_down_appendix}
As mentioned in Section~\ref{q_manage_and_async_locks}, a \texttt{lw\_rwlock} also supports upgrade and downgrade operations. Figure~\ref{rwlock_updown_algos} shows the algorithms for the two operations. Note that even though multiple readers could be traversing the \texttt{waitq} during upgrade, the traversal is safe. The \texttt{waitq} changes only due to arrival of new waiters to the back of the queue or removal of the waiter at the front of the queue during lock transfer. The former is immaterial to the traversal as it does not care for what happens to waiters behind it. The latter cannot happen as the traversing thread still has a read lock. The only possible race happens on the \texttt{next} field of the oldest waiter in the \texttt{waitq}: a reader performing an upgrade wants to add it's own waiter in front of it and a thread releasing write lock on a reader-biased lock is re-inserting list of existing waiters. This situation is handled by the upgrade logic and by the unlock routine. 

The unlock operation presented in Figure \ref{rwlock_algos} has 
to be slightly changed to support downgrade and upgrade of a \texttt{lw\_rwlock}.
For the case where a writer is releasing the write lock of a read-biased \texttt{lw\_rwlock},
while re-inserting the \texttt{wr\_q} at the front of the \texttt{waitq} of the \texttt{lw\_rwlock}, we have 
to use CAS instruction to co-ordinate with a possible upgrader. Also, if an upgrader is
found to be already present at the front of the \texttt{waitq}, the re-inserted \texttt{wr\_q} is added behind the upgrader's waiter.

\begin{figure}
\begin{center}
\begin{lstlisting}[multicols=2]
void downgrade(lw_rwlock_t $\ell$) {
  do {
    @$o$@ = @$n$@ = @$\ell$@;
    if (@$o$@.waitq != @NULLID@) {
      @{\em // Have existing waiters. Can't}@
      @{\em // do direct downgrade.}@
      break;
    }
    @$n$@.wlocked = 0;
    @$n$@.readers = 1;
  } while (!CAS(@$\ell$@, @$o$@, @$n$@));                              
  if (@$n$@.readers != 1) {    
    @$w$@ = tls_get_waiter();
    @{\em // Indicate that $w$ will now wait}@
    @{\em // for a read lock.}@
    @$w$@.app_data = @SHARED@;
    @insert\_waiter\_at\_front($\ell$.waitq, $w$)@;
    @{\em // Unlock will grant reader lock and}@
    @{\em // waiter will get a pending signal.}@
    unlock(@$\ell$@);
    @{\em // Consuming the pending signal}@
    @$w$@.event.wait();
  }
}

bool_t @upgrade(lw\_rwlock\_t $\ell$)@ {
  do {
    @$o$@ = @$n$@ = @$\ell$@;
    if (@$o$@.waitq != @NULLID@) {
      @{\em // There are more waiters that}@
      @{\em // could be upgrading themselves.}@
      break;
    } else if(@$o$@.readers == 1) {
      @{\em // Only reader, grab it right away.}@
      @$n$@.wlocked = 1;
      @$n$@.readers = 0;
    } else {
      @$w$@.app_data = @UPGRADE@;
      @$n$@.waitq = @$w$@.id; 
      @$n$@.readers --;
    }
  } while (!CAS(@$\ell$@, @$o$@, @$n$@));                              
  
  if (@$n$@.wlocked != 1) {
    @$w$@ = tls_get_waiter();
    if (@$n$@.waitq != @$w$@.id) {
      if(!@insert\_for\_upgrade($\ell$, $w$))@ {
        return @FALSE@; @{\em // failure}@ 
      }
      @{\em // Unlock and wait for lock}@
      @{\em // to be granted.}@ 
      unlock(@$\ell$@);
    }
    @$w$@.event.wait();
  }
  return @TRUE@; @{\em // success}@
}

bool_t @insert\_for\_upgrade(lw\_rwlock\_t $\ell$,@ 
                       @waiter\_t $w$)@ {
  while (@TRUE@) {
    @$ow$@ = find_oldest_waiter(@$\ell$@);
    if @($ow$.app\_data == UPGRADE)@ {
      @{\em // Someone else waiting for upgrade}@
      return @FALSE@;
    }
    @{\em // Try setting next pointer of last waiter.}@
    @{\em // Set the app\_data first since it needs to}@
    @{\em // be visible to any competing thread that}@
    @{\em // is also trying the upgrade.}@
    @$w$@.app_data = @UPGRADE@;
    if @(!CAS($ow$.next, NULLID, $w$.id))@ {
      @{\em // Competing upgrade or re-insert}@
      @$ow$@ = id2waiter(@$ow$@.next);   
      if @($ow$.app\_data == UPGRADE)@ {
        @{\em // Lost to competing upgrade}@
        return @FALSE@; @{\em // failure}@
      } @{\em // else lost to competing re-insert}@
    } else return @TRUE@; @{\em // CAS success}@
  }
}

\end{lstlisting}
\end{center}
\caption{Operations to downgrade and upgrade a \texttt{lw\_rwlock}.
The old and new values passed in to CAS are denoted by $o$ and $n$,
respectively. The caller's thread local waiter structure is denoted by $w$.
We use $ow$ to denote oldest waiter on the \texttt{waitq} (the head of the queue).}
\label{rwlock_updown_algos}
\end{figure}